\documentstyle[prl,aps]{revtex}
\begin{document}
\draft
\twocolumn[
\hsize\textwidth\columnwidth\hsize\csname@twocolumnfalse\endcsname
\title{Quasiparticle Resonant States Induced by a
Unitary Impurity in a $d$-Wave Superconductor}
\author{Jian-Xin Zhu,$^{(1)}$ T. K. Lee,$^{(2,3)}$
C. S. Ting,$^{(1,3)}$ and Chia-Ren Hu$^{(4)}$
}
\address{ 
 $^{(1)}$Texas Center for Superconductivity and Department of Physics, 
 University of Houston, Houston, Texas 77204\\
 $^{(2)}$ Institute of Physics, Academia Sinica, Nankang, 
Taipei 11529, Taiwan, R.O. China\\
 $^{(3)}$ National Center for Theoretical Sciences, P.O. Box 2-131,
 Hsinchu, Taiwan 300, R.O. China\\
 $^{(4)}$ Department of Physics,
 Texas A\& M University, College Station, Texas 77843
}
\maketitle
\begin{abstract}
The quasiparticle resonant states around a single nonmagnetic 
impurity with unitary scattering in a $d$-wave superconductor is 
studied by solving the Bogoliubov-de Gennes equations based on a 
$t$-$J$ model. Both the spatial variation of the order parameter 
and the local density of states (LDOS) around the impurity have 
been investigated. We find: (i) A particle-hole symmetric system 
has a single symmetric zero-energy peak in the LDOS regardless of 
the size of the superconducting coherence length $\xi_0$; (ii) For 
the particle-hole asymmetric case, an asymmetric splitting of the 
zero-energy peak is intrinsic to a system with a small value of
$k_{F}\xi_0$.
 
\end{abstract} 
]

\narrowtext
It is now well established~\cite{Scalap95} that high-$T_c$
superconductors (HTSC's) have essentially a $d_{x^2-y^2}$-wave 
pairing symmetry. In conventional $s$-wave superconductors, 
nonmagnetic impurities affect neither the transition temperature 
nor the superfluid density as dictated by the Anderson 
theorem~\cite{Anderson59}. But in a $d$-wave superconductor 
(DWSC) with nodes of the energy gap, such impurities can 
cause a strong pair-breaking effect~\cite{Gorkov84}. Recently, 
the local electronic properties in the immediate vicinity of an 
isolated non-magnetic impurity in a DWSC has become the topic 
of increased investigation~\cite{Lee93,BFS93,Choi94,BSR95,SBS96,%
FKB96,OOSM96,FB97,YHLK99,HPGND99,SAGY99,TKS98,TTOK99}, as 
these properties may provide a distinctive signature for the pairing 
symmetry. It has been theoretically predicted by Balatsky, Salkola 
and co-workers~\cite{BSR95,SBS96} that, in a DWSC, a single 
nonmagnetic impurity can generate quasiparticle resonant 
states at subgap energies. They showed that, for a moderately 
strong impurity, an asymmetry of the resonance peak near the Fermi 
energy is induced by the fact that the impurity locally breaks the 
particle-hole symmetry. However, their theory says that increasing
the impurity strength pushes the resonance peak toward the Fermi 
level, so that, in the unitary limit, the resonance occurs right 
on the Fermi level, and only a single symmetric zero-energy peak 
(ZEP) occurs in the LDOS near the impurity. It has also been shown 
by a finite-size diagonalization~\cite{OOSM96} that, in the unitary 
limit, the lowest eigenvalues are essentially zero, indicative of 
the appearance of zero-energy states (ZES's). Note that, 
in Ref.~\cite{OOSM96}, the chemical potential $\mu$ was taken to be 
at the center of the tight-binding energy band (i.e., $\mu = 0$), 
so that the system has a particle-hole symmetry. This symmetry is 
also upheld in the continuum-theory treatment 
of impurities~\cite{BSR95,SBS96} where the self-consistent $t$-matrix 
approximation is employed. A question which arises naturally is 
whether, in the unitary limit, the ``ZEP'' in the LDOS due to the 
``ZES's'' has an asymmetric splitting or not, when particle-hole 
symmetry is broken in the system. Recently, Tanaka 
{\it et al.}~\cite{TKS98} concluded with their numerical study 
that such a splitting is still present, whereas 
Tsuchiura {\it et al.}~\cite{TTOK99} made an opposite conclusion 
in their numerical study, and asserted that the system studied by 
Tanaka {\it et al.} was too small for their results to be reliable. 
Experimentally, an asymmetric splitting is clearly observed by 
Yazdani {\it et al.}~\cite{YHLK99}, whereas Hudson 
{\it et al.}~\cite{HPGND99} observed only an off-centered 
peak with no indication of a splitting. Thus it appears 
important to settle the issue of whether a unitary non-magnetic 
impurity in a pure DWSC can indeed give rise to such an asymmetric 
splitting in the ``ZEP'', as it will decide whether experimental 
observation of this feature in HTSC's necessarily implies that 
these SC's do not have pure $d$-wave symmetry, or that the 
impurity is not in the unitary limit (in which case the asymmetry 
is tied to the {\it sign} of the impurity potential, which may 
well be a misleading conclusion).  
 
Based on a $t$-$J$ model, this paper presents an extensive study 
on the electronic states around a unitary single-site impurity
in a DWSC. The spatial variation of the superconducting order 
parameter (OP) near the impurity, including an induced $s$-wave
component, is determined self-consistently. By investigating the 
sensitivity of the LDOS on both $\mu$ and $\xi_0$, we find: 
(i) When $\mu=0$, so that the system is particle-hole symmetric, 
a single ZEP occurs in the LDOS spectrum which is symmetric with 
respect to zero energy, regardless of the size of $\xi_0$; 
(ii) As the particle-hole symmetry is broken by letting $\mu 
\neq 0$, a critical value $\gamma_c$ exists, which is larger for 
larger $\vert\mu\vert$, so that for $\gamma \equiv k_{F}\xi_0 
< \gamma_c$ the ``ZEP'' exhibits an asymmetric splitting. 
(Here $k_{F}$ is the Fermi wavevector.)~\cite{joynt97} 
Thus we find that for a particle-hole asymmetric system, a 
sufficiently small coherence length can cause the ``ZEP'' to 
exhibit an asymmetric splitting. Treating such a system by 
the self-consistent $t$-matrix approximation, which restores 
the particle-hole symmetry, will then lose this feature and be 
misleading in this respect.

We consider a $t$-$J$ model Hamiltonian defined on a 
two-dimensional square lattice: 
\begin{eqnarray}
{\cal H} &=& -t \sum_{\langle{\bf ij}\rangle \sigma}  
c_{{\bf i}\sigma}^{\dagger}c_{{\bf j}\sigma}
+\sum_{{\bf i}}U_{\bf i}n_{{\bf i}}
-\mu \sum_{{\bf i}}n_{{\bf i}} \nonumber \\
&&+\frac{J}{2}
\sum_{\langle {\bf ij}\rangle }[
{\bf S}_{{\bf i}}\cdot {\bf S}_{{\bf j}}-\frac{1}{4}
n_{\bf i}n_{\bf j}]
+\frac{W}{2}\sum_{\langle {\bf ij}\rangle}n_{\bf i}n_{\bf j}\;,
\label{EQ:t-J}
\end{eqnarray}
where the Hilbert space is made of empty and singly-occupied sites
only; summing over $\langle {\bf ij}\rangle$ means summing 
over nearest-neighbor sites; $n_{\bf i} = \sum_{\sigma}c_{{\bf 
i}\sigma}^{\dagger} c_{{\bf i}\sigma}$ is the electron number 
operator on site ${\bf i}$; 
${\bf S}_{\bf i}$ is the spin-$\frac{1}{2}$ operator on site 
${\bf i}$; and $J > 0$ gives the antiferromagnetic superexchange 
interaction.  As in Ref.~\cite{MRR90}, we have also included a 
direct nearest-neighbor interaction term. $W = 0$ and $J/4$ 
correspond to two versions of the standard $t$-$J$ model. 
This term is introduced to adjust the magnitude of the 
resultant $d$-wave OP.  The 
scattering potential from the single-site impurity is modeled by 
$U_{\bf i} = U_{0}\delta_{{\bf i}I}$ with $I$ the index for the 
impurity site. The slave-boson method~\cite{KL88} is employed to 
write the electron operator as $c_{{\bf i}\sigma} = b_{\bf
i}^{\dagger}f_{{\bf i}\sigma}$, where $f_{{\bf i}\sigma}$ and 
$b_{\bf i}$ are the operators for a spinon (a neutral 
spin-$\frac{1}{2}$ fermion) and a holon (a spinless charged boson). 
Due to the holon Bose condensation at low temperatures, the 
quasiparticles are determined by the spinon degree of freedom 
only. Within the mean-field approximation, the Bogoliubov-de 
Gennes (BdG) equations are derived to be
\begin{equation}
\label{EQ:BdG}
\sum_{\bf j} \left( \begin{array}{cc} 
H_{\bf ij} & \Delta_{\bf ij} \\
\Delta_{\bf ij}^{\dagger} & -H_{\bf ij} 
\end{array} \right) \left( \begin{array}{c}
u_{\bf j}^{n} \\ v_{\bf j}^{n} 
\end{array} \right)
= E_{n} \left( \begin{array}{c}
u_{\bf i}^{n} \\ v_{\bf i}^{n} 
\end{array} \right)\;,
\end{equation}
with
\begin{equation}
H_{\bf ij} = -[t\delta+(\frac{J}{2}+W)\chi_{\bf ij}]
\delta_{{\bf i}+\mbox{\boldmath{$\delta$}},{\bf j}}
+(U_{\bf i}-\mu)\delta_{\bf ij}\;.
\end{equation} 
Here $u_{\bf i}^{n}$ and $v_{\bf i}^{n}$ are the Bogoliubov 
amplitudes corresponding to the eigenvalue $E_{n}$;
$\delta$ and $\chi_{\bf ij}$ are the doping rate and the 
bond OP, respectively; and $\mbox{\boldmath{$\delta$}}$ are
the unit vectors $\pm \hat{\bf x}$, $\pm \hat{\bf y}$. 
The resonant-valence-bond (RVB) OP 
$\Delta_{\bf ij}$, $\chi_{\bf ij}$, and $\delta$ 
are determined self-consistently: 
\begin{equation}
\Delta_{\bf ij}= \frac{J-W}{2}\sum_{n}
[u_{\bf i}^{n}v_{{\bf j}}^{n*}
+u_{{\bf j}}^{n}v_{\bf i}^{n*}]
\tanh (E_{n}/2k_{B}T) 
 \delta_{{\bf i}+\mbox{\boldmath{$\delta$}},{\bf j}}\;,
\end{equation}
\begin{equation}
\chi_{\bf ij} = \sum_{n}\{ u_{\bf i}^{n*}u_{\bf j}^{n}f(E_{n})
+v_{\bf i}^{n}v_{\bf j}^{n*}[1-f(E_n)]\}\;,
\end{equation}
and
\begin{equation}
\delta = 
1-\frac{2}{N_a}\sum_{{\bf i},n}\{\vert u_{\bf i}^{n}\vert^{2}
f(E_n)+\vert v_{\bf i}^{n} \vert^{2} [1-f(E_n)]\}\;,
\end{equation}
where $k_{B}$ is the Boltzmann constant; $f(E) = 
[\exp(E/k_{B}T)+1]^{-1}$ is the Fermi distribution function; 
and $N_a = N_x\times N_y$ is the number of lattice sites. 
The BdG equations is solved fully self-consistently 
for the bulk state first. We then fix the values of $\delta$ and $\chi$ 
and solve the BdG equations in the presence of a single 
impurity with the self-consistent $d$-wave RVB OP.  
The thermally broadened local density of states (LDOS) is 
then evaluated according to:
\begin{equation}
\rho_{\bf i}(E) = -2\sum_{n}[\vert 
u_{\bf i}^{n}\vert^{2} 
f^{\prime}(E_{n}-E) +\vert v_{\bf 
i}^{n}\vert^{2}f^{\prime}(E_{n}+E)]\;, 
\end{equation}
where a factor $2$ arises from the spin sum, and $f^{\prime}(E) 
\equiv df(E)/dE$. The LDOS $\rho_{\bf i}(E)$ is proportional to 
the local differential tunneling conductance which can be 
measured in a scanning tunneling microscope/spectroscopy 
(STM/S) experiment~\cite{Tinkham75}.

In the numerical calculation, we construct a superlattice with 
the square lattice $N_x\times N_y$ as a unit supercell. As 
detailed in Ref.~\cite{ZFT99}, this method can provide the 
required energy resolution for the possible resonant states.  
Throughout the work, we take the size of the unit supercell  
$N_a=35\times 35$, the number of supercells $N_c=6\times 6$, 
the temperature $T=0.01J$, and the single impurity potential 
in the unitary limit $U_0=100J$. The values of the other 
parameters --- $\mu$, $W$, and $t$, are varied in order to 
investigate the electronic states around a single impurity 
for various ways to bring about particle-hole asymmetry. 
The obtained spatial variation of the $d$-wave 
and the induced extended-$s$-wave OP components around the 
impurity, which are defined as 
$\Delta_{d}({\bf i})=\frac{1}{4}[\Delta_{\hat{x}}({\bf i})
+\Delta_{-\hat{x}}({\bf i})
-\Delta_{\hat{y}}({\bf i})
-\Delta_{-\hat{y}}({\bf i})]$,  
and $\Delta_{s}({\bf i})=\frac{1}{4}[\Delta_{\hat{x}}({\bf i})+
\Delta_{-\hat{x}}({\bf i})
+\Delta_{\hat{y}}({\bf i})
+\Delta_{-\hat{y}}({\bf i})]$, 
is similar to Fig. 1 of Ref.~\cite{FKB96}. These OP component's 
have the following characteristics: The $d$-wave component decreases 
continuously to zero from its bulk value as the impurity site 
is approached, in the scale of the coherence length $\xi_0 \equiv 
\hbar v_{F}/\pi\Delta_{max}$, with the depleted region extending 
farther in the nodal directions if $\xi_0 $ is larger. 
Here $\Delta_{max} = 4\Delta_0$ with $\Delta_0$ the bulk value 
of the $d$-wave OP defined in the real space on a nearest neighbor 
bond, and $v_{F}$ is the Fermi velocity. The $s$-wave component 
is zero at the impurity site and also at infinity. It has line-nodes 
along the $\{110\}$ and $\{1\bar{1}0\}$ directions, and changes sign 
across any nodal line. Unlike the pairing state at a $\{110\}$ 
surface of a DWSC, which can break the time-reversal symmetry, 
the pairing state near a single impurity conserves time-reversal 
symmetry. This difference can be understood from the 
Ginzburg-Landau (GL) theory~\cite{XRT95}, in that a mixed gradient 
term favors the $d$- and induced $s$-wave OP components to be in 
phase, but it vanishes near a $\{110\}$ surface, whence the fourth 
order $s$-$d$ coupling term can establish an $s+id$ pairing state.
  
Figure~\ref{FIG:LDOS} shows the LDOS as a function of energy on sites 
one and two lattice spacings along the $(100)$ direction from the 
impurity and on the corner site of the unit supercell. The values of 
the parameters are labeled on the figure. Note that the LDOS at the 
corner site has recovered the bulk DOS, by exhibiting a gaplike 
feature with the gap edges at $\pm\Delta_{max}$. This resemblance 
indicates that the unit cell size and the number of unit cells 
are large enough for uncovering the physics intrinsic to an 
isolated impurity.  As shown in Fig.~\ref{FIG:LDOS}, we find that 
the LDOS spectrum near the impurity is highly sensitive to the 
position of $\mu$ within the energy band, and 
the parameter $\gamma$. In Fig.~\ref{FIG:LDOS}(a), $\mu=0$~\cite{note} and 
$\gamma = 0.80$, a single ZEP occurs in the LDOS on the 
nearest-neighbor site of the impurity, similar to the prediction 
of the continuum theory~\cite{BSR95,SBS96} and the eigenvalue 
calculation in Ref.~\cite{OOSM96}. In addition, as a reflection 
of the particle-hole symmetry, the whole LDOS spectrum is 
symmetric about $E=0$. We have also studied the cases (not shown) 
with the same $\mu=0$ and $t=4J$ but with $W = 0$ and $W=0.5J$ 
(corresponding to $\gamma = 0.27$ and $2.0$), and found that 
the above feature remains unchanged, which allows us to 
conclude that as long as the system is particle-hole symmetric, 
only a single symmetric ZEP exists for all $\gamma$. When $\mu$ 
is not zero, the system is particle-hole asymmetric, and the 
LDOS spectrum becomes asymmetric. 
[See Fig.~\ref{FIG:LDOS}(b)-(g).] In Fig.~\ref{FIG:LDOS}(b)-(e), 
$\mu = -0.32J$ is fixed, and $W$ and $t$ are varied in order to 
change $\gamma$. For a large $\gamma = 91.7$, we see a 
single ZEP in the LDOS. [See Fig.~\ref{FIG:LDOS}(b).] When $\gamma$ 
is lowered to $16.5$, the ``ZEP'' begins to evolve into a 
double-peaked structure with the $E>0$ peak having the dominant 
spectral weight over the $E<0$ peak.  For a further decreased 
$\gamma = 6.05$, the spectral weight of the peak at $E<0$ is 
enhanced. (See Fig.~\ref{FIG:LDOS}(d).) As seen in 
Fig.~\ref{FIG:LDOS}(e), this enhancement becomes even more 
pronounced when $\mu$ is made close to the edge of a very 
narrow energy band so that $\gamma$ becomes as small as $2.85$. 
When $\mu = -0.16J$, we only observe a single ZEP although 
$\gamma$ is as small as 5.7 [for Fig.~\ref{FIG:LDOS}(f)] and $2.9$ 
[for Fig.~\ref{FIG:LDOS}(g)], except that a tendency of the 
splitting can be identified in the latter case. 
This tendency of the splitting has been observed clearly in STM tunneling 
spectroscopy measurements (see Fig.~4(A) of Ref.~\cite{YHLK99}). It 
should be emphasized 
that the ZEP splitting obtained here has a different origin from that 
found by Tanaka et al.~\cite{TKS98}. We have re-examined their 
results by choosing the same parameter values and the system 
size ($18\times 18$). When the LDOS spectrum is displayed in a wide 
energy landscape, many split DOS peaks appear with no well-defined 
gaplike feature identifiable. But as the system size is enlarged by 
the supercell technique, the calculation only shows a single ZEP in 
the LDOS, which indicates that the splitting of ZEP obtained in 
Ref.~\cite{TKS98} is indeed due to the size effect.  On the other 
hand, we have also calculated the excess charge distribution 
due to the presence of the impurity ($\propto\delta n_{\bf i} = 
\langle n_{\bf i}\rangle -n_0$, where $n_0$ is the average particle 
occupation on each site for the bulk system). We find that this 
distribution is anisotropic, with its magnitude having tails 
extending along the nodal directions [See Fig.~\ref{FIG:DENSITY}]. 
Because Fig.~\ref{FIG:DENSITY} is obtained with the parameter 
values given in Fig.~\ref{FIG:LDOS}(e) which gives a small 
$\gamma\;(= 2.85)$ value, the exhibited tail is short. 
A similar calculation with the model parameters given in 
Fig.~\ref{FIG:LDOS}(f) (not shown) shows that the charge 
distribution is similar to that displayed in 
Fig.~\ref{FIG:DENSITY} except for a longer tail along 
the nodal directions due to the larger $\gamma\;(= 5.7)$. 
This similarity in the charge distributions for a split and a 
unsplit ZEP's disproves the assertion made in Ref.~\cite{TKS98} 
that the local charge-density oscillation is the cause of 
the ZEP splitting. We mention in passing that we have 
also found that the excess charge density decays exponentially 
along the nodal directions instead of the $r^{-2}$-%
dependence from the impurity. But we do not think 
that this finding invalidates the assertion in Ref.~\cite{BSR95}
that the wavefunction of the impurity resonant state has a 
$1/r$ decay along the nodal directions, which can lead to a 
long range interaction between the impurities. However, 
we do believe that since we have obtained essentially the 
bulk density of states in several neighboring points near the 
corner of the supercell, the interaction between the neighboring 
impurities should be negligible in the cell size we have chosen 
to work with. Thus we believe that it is very unlikely that the 
splitting of the ZEP we obtain is due to this interaction. 
Since the $s$-wave OP component induced near the impurity is 
in phase with the dominant $d$-wave component, the splitting of 
the ZEP we found is not due to a local broken time-reversal 
symmetry. Finally, as shown in Fig.~\ref{FIG:LDOS}(e), 
the splitting is also exhibited in a non-self-consistent 
calculation with a spatially uniform bulk $d$-wave OP, showing 
that the suppression of the $d$-wave OP component, and the induction 
of the $s$-wave component, have little to do with the splitting.  
All of these points lead us to the conclusion that, for the 
particle-hole asymmetric case, the splitting of the ZEP is
intrinsic to the system with a short coherence length, and the 
critical value $\gamma_c$, below which the ZEP is split into an 
asymmetric double-peak, is simply a reflection that the system 
has reached a critical extent in its deviation from particle-hole 
symmetry. We thus propose to understand these results qualitatively
as follows: The ``ZES's'' induced by a unitary non-magnetic impurity  
have essentially the same physical origin as the ``midgap states'' 
predicted to exist on the surfaces/interfaces of a 
DWSC~\cite{hu94}. Their existence is implied topologically by 
the Atiyah-Singer index theorem~\cite{indexthm}, which applies to 
particle-hole-symmetric Dirac-like operators. When this symmetry 
is mildly broken the midgap states are expected to still exist but 
no longer exactly ``midgap''. The BdG equations become 
Dirac-like equations only under the WKBJ approximation (which is a 
part of the self-consistent $t$-matrix approximation), the error 
of which is measured by the 
parameters $\vert\mu\vert$ and $\gamma^{-1}$. 
For $\mu = 0$, the system has exact particle-hole symmetry for all $\gamma$. 
Thus smaller $\vert\mu\vert$ should 
imply smaller $\gamma$ needed to reach the same deviation 
from particle-hole symmetry, hence a smaller $\gamma_c$ below 
which an asymmetric splitting of the ZEP appears. 

In summary, we have presented an extensive study on the quasiparticle 
resonant states induced by a unitary non-magnetic impurity in a 
DWSC. The results have clarified some conflicting conclusions 
in the literature, and should be of value for the proper analysis 
of the STM/S results obtained on HTSC's around an isolated impurity.

We are grateful to M. Salkola, M. E. Flatt\'{e}, and A. Yazdani for 
valuable discussions. This work was supported by the Texas Center for 
Superconductivity at the University of Houston, the Robert A. Welch
Foundation, and the Texas Higher Education Coordinating Board. 
Free computing time from the Texas A\&M Supercomputer Center is also 
greatfully acknowledged.

\begin{figure}
\caption{Local density of states as a function of energy on sites 
one (solid line) and two lattice (dashed line) spacings along the 
$(100)$ direction away from the impurity, and on the corner site
(short-dashed line) of the unit cell. The parameter values have been
correspondingly labeled on each panel. Also shown in the (e) panel  
is the local density of states on the site nestest-neighbor to the 
impurity obtained with a pure bulk $d$-wave order parameter (dotted
line).} 
\label{FIG:LDOS} 
\end{figure}

\begin{figure}
\caption{Spatial variation of the charge distribution around the 
impurity with the parameter values given in Fig.~\ref{FIG:LDOS}(e).}
\label{FIG:DENSITY} 
\end{figure}


\begin{references}
\bibitem{note} Notice that at small doping, in particular at zero doping, 
the mean-field slave-boson approach adopted here is not representative of 
the physics of the $t$-$J$ model at all, but just a tool for analyzing the 
resonant states in the particle-hole symmetric limit.
\end{references}
\end{document}